\documentclass[12pt,3p]{article}

\usepackage{amssymb}
\usepackage{graphicx}
\usepackage{amsthm,amssymb,amsfonts,subfigure}
\usepackage[english]{babel}
\usepackage{amsmath}

\usepackage{epstopdf}
\usepackage[title]{appendix}
\usepackage{pdflscape}
\usepackage{algorithm}
\usepackage{algorithmic}
\newtheorem{definition}{Definition}

\usepackage[utf8x]{inputenc}

\usepackage{lineno}
\makeatletter
\def\makeLineNumberLeft{%
  \linenumberfont\llap{\hb@xt@\linenumberwidth{\LineNumber\hss}\hskip\linenumbersep}
  \hskip\columnwidth
  \rlap{\hskip\linenumbersep\hb@xt@\linenumberwidth{\hss\LineNumber}}\hss}
\leftlinenumbers
\makeatother

\begin{document}

\title{Global Rank Estimation}

\author{%
  Akrati Saxena\\akrati.saxena@iitrpr.ac.in
  \and S. R. S. Iyengar\\sudarshan@iitrpr.ac.in
  \and Department of Computer Science and Engineering,\\ Indian Institute of Technology Ropar, India
  }
\date{
}

\maketitle


\begin{abstract}

In real world complex networks, the importance of a node depends on two important parameters: 1. characteristics of the node, and 2. the context of the given application. The current literature contains several centrality measures that have been defined to measure the importance of a node based on the given application requirements. These centrality measures assign a centrality value to each node that denotes its importance index. But in real life applications, we are more interested in the relative importance of the node that can be measured using its centrality rank based on the given centrality measure. To compute the centrality rank of a node, we need to compute the centrality value of all the nodes and compare them to get the rank. This process requires the entire network. So, it is not feasible for real-life applications due to the large size and dynamic nature of real world networks.

In the present project, we aim to propose fast and efficient methods to estimate the global centrality rank of a node without computing the centrality value of all nodes. These methods can be further extended to estimate the rank without having the entire network. The proposed methods use the structural behavior of centrality measures, sampling techniques, or the machine learning models. In this work, we also discuss how to apply these methods for degree and closeness centrality rank estimation.

\end{abstract}

\section{Introduction}


In complex networks, each node has some unique characteristics that are used to decide its importance based on the application context. For example, in a social network, a node having the high degree will be more important if the nodes are selected for an online social polling\footnote{In online social polling, a user vote based on the behavior of its neighbors \cite{dasgupta2012social}. For example, if a company surveys that most of the people are using Microsoft office or Open office, then the user will vote for Microsoft office if most of her friends are using this otherwise for open office. It can also be weighted where a user can specify what percentage of people are using which software.} but an influential node is preferred over high degree node if it is required to spread some information \cite{kitsak2010identification}. Similarly, if the government wants to set up a new service like a hospital, school etc. in a city, then a place having high closeness with all other places will be chosen but if the service is like setting up a checking booth then a point that connects more places will be a better choice as more people will pass through it. These examples help us to understand that the importance of a node changes based on the requirement.

In real life applications, we are interested in selecting few top important nodes based on the context as we have limited resources and they need to be used optimally. The efficient methods are required to identify these important nodes faster in the given network. The current literature contains various centrality metrics to measure the importance of a node based on the given application context. These centrality measures can be categorized as local centrality measures and global centrality measures. Centrality measures that can be computed using local/neighborhood information of the node are called local centrality measures like degree centrality \cite{shaw1954some}, H-index \cite{hirsch2005index}, semi-local centrality \cite{chen2012identifying} etc. The centrality measures that require the entire network for their computation are called global centrality measures like closeness centrality \cite{sabidussi1966centrality}, betweenness centrality \cite{freeman1977set}, eigenvector centrality \cite{stephenson1989rethinking}, coreness \cite{seidman1983network}, PageRank \cite{brin1998anatomy} etc. The computational complexity of global centrality measures is very high and it depends on the network size.

\subsection{Definitions}

In this section, we discuss the definition of all basic centrality measures. The terminologies used in the paper are explained here. A graph is represented as $G(V,E)$, where $V$ is the set of nodes and $E$ is the set of edges. $n$ represents total number of nodes and $m$ represents total number of edges in the network. $d_u$ represents degree of node $u$.  

\begin{definition}{\textbf{Degree Centrality:}}
Degree Centrality of a node $u$ is defined as,
\begin{center}
$C_D(u)= \frac{d_u}{n-1}$
\end{center}
\end{definition}

\begin{definition}{\textbf{Closeness Centrality:}}
Closeness centrality represents the closeness of a given node with every other node of the network. In precise terms, it is inverse of the farness which in turn is the sum of distances with all other nodes. The closeness centrality \cite{freeman1978centrality} of a node $u$ is defined as,
\begin{center}
$C_C(u)= \frac{n-1}{\sum_{\forall v, v \neq u}d(u,v)}$
\end{center}
\end{definition}
In disconnected graphs, the distance between all pairs of nodes is not defined, so this definition of closeness centrality can not be applied for disconnected graphs.

\begin{definition}{\textbf{Betweenness Centrality:}}
Betweenness centrality of a given node $u$ is based on the number of shortest paths passing through the node \cite{freeman1978centrality}. This measure basically quantifies the number of times a node acts as a bridge along the shortest path between a pair of nodes. The betweenness centrality of a node $u$ is defined as,
\begin{center}
$C_B(u)= \frac{\sum_{s \neq u \neq t}\frac{\partial_{st}(u)}{\partial_{st}}}{(n-1)(n-2)/2}$
\end{center}
where $\partial_{st}(u)$ represents the number of shortest paths between nodes $s$ and $t$ with node $u$ acting as an intermediate node in the shortest path, and $\partial_{st}$ represents total number of shortest paths between nodes $s$ and $t$.

\end{definition}

\begin{definition}{\textbf{Eigenvector Centrality:}}
Eigenvector Centrality is used to measure the influence of a node in the network \cite{stephenson1989rethinking}. It assigns a relative index value to all nodes in the network based on the concept that connections with high indexed nodes contribute more to the score of the node than the connections with low indexed nodes.

The Eigenvector centrality for a graph $G(V,E)$ is given as,
\begin{center}
$C_E(u)= (1/\lambda) \sum A_{uv} C_E(v)$
\end{center}

where $A$ is the adjacency matrix of the graph, $v$ is the neighbour of $u$ and $\lambda$ is a constant.
\end{definition}


\begin{definition}{\textbf{Katz Centrality:}}
Katz centrality was introduced by Katz in 1953 to measure the influence of a node \cite{katz1953new}. It assigns different weights to shortest paths according to their lengths, as the shorter paths are more important for information flow than the longer paths. Contribution of a path of length $P$ is directly proportional to $s^P$ and $s \in (0,1)$. It is defined as,
\begin{center}
$K= sA + s^2A^2 + s^3A^3 + ... + s^PA^P + .... = (I-sA)^{-1} -I$
\end{center}

where $I$ is a unit matrix, $A$ is the adjacency matrix of the graph.
\end{definition}

%

\subsection{Related Literature}

Centrality measures have attracted research community due to their importance in complex networks. The comparison of relative importance of the nodes is required in many real life applications like ranking the web pages in web search, identifying the most influential nodes in a social network, etc. The state of the art literature on centrality measures can be divided into the following main categories.

\begin{enumerate}
\item \textbf{Extensions of Centrality Measures:} In 1977, Freeman proposed three main centrality measures to identify the importance of nodes based on the local and global connectivity \cite{freeman1977set}. The proposed definitions were applicable for undirected and unweighted networks. But these unweighted networks are not enough to convey the complete information of the system. Many complex systems are represented using the variety of networks which require redefining the centrality metrics. So, the centrality measures have been extended for different types of networks like directed networks \cite{du2015new}, weighted networks \cite{opsahl2010node}, multiplex networks \cite{barzinpour2014clustering}, disconnected networks \cite{latora2001efficient}, and so on. 

\item \textbf{Approximation Methods:}
The computation cost of global centrality measures is very high in large scale networks. It has motivated researchers to propose fast and efficient approximation methods for global centrality measures \cite{agarwal2014bolt, eppstein2004fast, dellingcomputing, rattigan2006using, shi2011fast, eppstein2001fast, newman2005measure, bergamini2015fully, geisberger2008better, riondato2014fast, amento2000does}. These approximation methods can be efficiently used to compare two nodes, where we don't need actual centrality values.

\item \textbf{Update Centrality Values in Dynamic Networks:} In large scale dynamic networks, it is not feasible to recompute centrality values for each network update due to their high computational complexity. So, researchers have proposed various methods to update centrality values in dynamic networks \cite{singh2015faster, lee2012qube, green2012fast, yu2013irwr, kas2013incremental, yen2013efficient, brandes2001faster, sariyuce2013incremental}.

\item \textbf{Identification of top-k nodes:} In many applications, we are only interested in identifying top few nodes. For example, select top-$k$ people to distribute free samples of a product or top-$k$ nodes in the Internet system for the robust instant backup, etc. The computation of centrality values of all nodes and identification of top-k nodes will be very costly. There are methods to directly identify top-k nodes in the network without computing the centrality of all nodes \cite{ufimtsev2014extremely, okamoto2008ranking, olsen2014efficient, bergaminicomputing, avrachenkov2010monte, cao2010retrieving}. It reduces the time complexity drastically.

\item \textbf{Applications of Centrality Measures:} Centrality measures have been applied to different types of networks and their performance is studied to rank the nodes. For example,  Newman used closeness and betweenness centrality to better understand collaboration networks \cite{newman2001scientific}. He shows that most of the shortest paths of a node pass through only top few collaborators and remaining collaborators participate in very less number of shortest paths. Yan et al. also studied various parameters of collaboration networks using closeness centrality \cite{yan2009applying}. Sporns et al. used closeness centrality to identify hubs in the brain network \cite{sporns2007identification}. Iyengar et al. showed that the closeness centrality plays an important role in human navigation \cite{sudarshan2012understanding}. Leydesdorff used betweenness centrality to study the citation network of Journals \cite{leydesdorff2007betweenness}. They show that betweenness centrality is a good parameter to measure the interdisciplinarity of journals using local citation environments. Cheng et al. used pagerank and HITS to rank nodes in Journal citation networks and show that and it gives good ranking than ISI impact factor \cite{cheng2009pagerank}.

\item \textbf{Others:} A few other related works are like the computation of centrality measures for large size distributed networks \cite{wang2015distributed, jakma2012distributed}, parallel algorithms to compute the centrality values \cite{bader2006parallel}, hybrid centrality measures \cite{abbasi2013hybrid, basaras2013detecting}, etc. Hybrid centrality measures are the combination of two or more centrality measures and are used for the specific types of networks. Researchers have also studied the correlation of these centrality measures to better understand the structure and evolution of complex networks \cite{valente2008correlated, tallberg2000comparing}. Evolving models proposed using centrality based analysis include \cite{ko2008rethinking, saxena2016evolving}.

\end{enumerate}

\begin{figure}[]
\centering
\includegraphics[width=14cm]{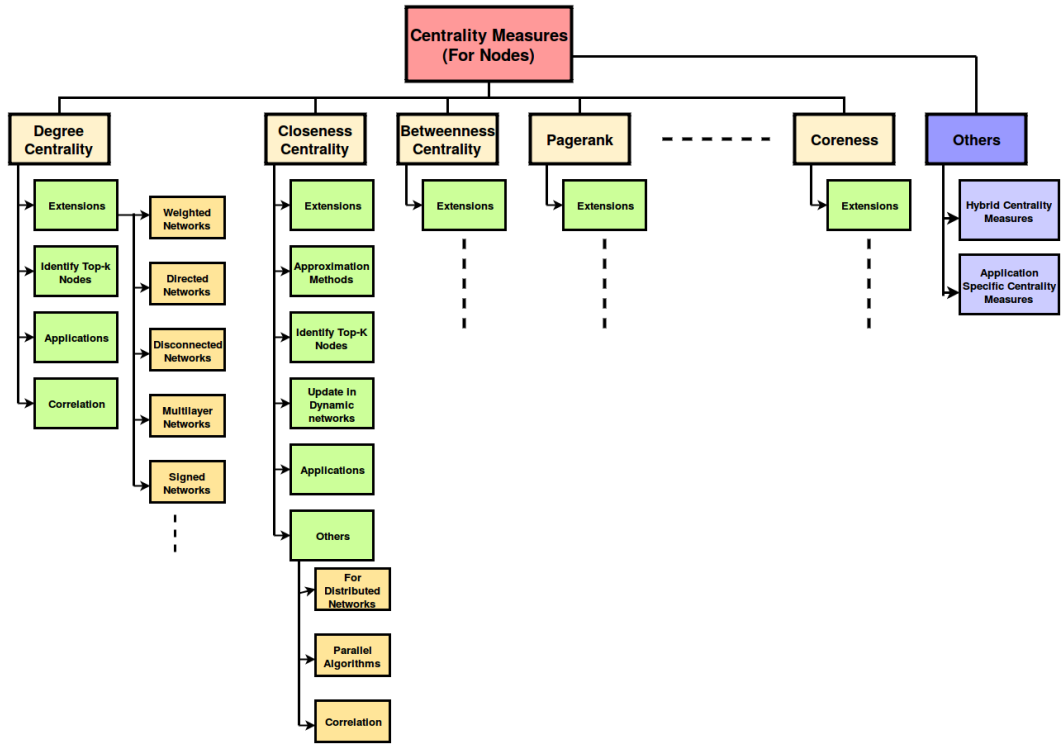}
\caption{Categorization of research work on Centrality Measures}
\label{categorization}
\end{figure}

The brief categorization of research work on centrality measures is shown in Figure~\ref{categorization}.

Centrality measures also have been defined for a group of nodes to measure how central a group is with respect to the given network \cite{everett1999centrality, ni2011degree, zhao2014measuring, garrido2016group, aleskerov2016centrality, chen2016efficient, jiang2011using, tejada2014game, puzis2008group, flores2014game}. For example, closeness centrality of a group of nodes is computed to understand how close these nodes are in the given network. Similarly, coreness of a group of nodes is computed to measure how tightly knit these nodes are with each other and with rest of the network.
The network centrality measures are used to compare networks based on their structural characteristics. The network centralities can be measured using coreness index \cite{della2013profiling}, density, average path length, diameter, average clustering coefficient, and so on. The coreness centrality index \cite{della2013profiling} of a network denotes the extent of core-periphery structure \cite{borgatti2000models} in the given network. The metrics like density, average path length, diameter, average clustering coefficient, consider the connectivity pattern of nodes in the given network.

\subsection{Motivation}

The main objective to define centrality measures is to rank nodes in the given network. Currently, we follow a two-step process to compute the rank, 1. compute centrality values of all the nodes, and 2. compare them to measure the rank of a node. The complexity of this classical method is very high as it includes the complexity to compute the centrality values of all nodes, and the complexity to compare them to find the rank of a node. This method also requires the entire network to measure the rank of a single node.

In our project, we aim to propose fast and efficient methods to directly estimate the rank based on the given centrality measure without computing the centrality values of all nodes, with less time and space complexity. We further aim to propose rank estimation methods based on the local information of the node without having the entire network.


\section{The Proposal}

The comparison of the relative importance of the nodes is required in many real-life applications. In a social network, a person might be interested to estimate her rank to know how strong she is in the network based on the given context. This problem can be explained by a simple phrase; most of the time, the actual value is not important, \textit{what's important is where you stand}, not with respect to the mean but with respect to others. In the current project, we focus on estimating the rank of a node based on the given centrality measure.

The centrality rank of a node $u$ is defined as, $R_{act}(u) = \sum_{v} X_{uv} + 1$, where $X_{uv} = 1,$ if $C(v) > C(u)$, otherwise $X_{uv} = 0$. $C(u)$ denotes the centrality value of the node $u$ based on the given centrality metric. It has been referred as actual centrality rank throughout the paper. A node having the highest centrality value is ranked $1$, and all nodes having the same centrality value will have the same rank. The node who is interested in computing its centrality rank is referred as \textit{Interested Node}. In this project, our aim is to propose fast and efficient methods to estimate the centrality rank of a node.

Real world complex networks are highly dynamic, so, the importance of the nodes keeps on changing. It is very costly to apply the exact or approximation algorithms of different centrality measures to calculate the centrality rank of a node. For example, to compute the closeness rank of a node, first, we compute closeness centrality of all nodes and then compare them to compute the rank of a node. The complexity to compute closeness centrality of all nodes is $O(n \cdot m)$ that is very high. Instead of this, the network properties can be exploited to propose methods having less complexity. This motivates to propose efficient methods to estimate the centrality rank of a node without computing the centrality values of all nodes. Except this, It is also not feasible to download the entire network and process it due to its large size. Thus, one can see the importance of predicting the global rank of a node using its local information without collecting the entire network. It will be a good future direction to understand the ranking pattern of different centrality measures and predict the rank with high accuracy. 

Next, we discuss the prospective research directions related to the project.

\begin{itemize}

\item \textbf{Propose methods having less complexity to estimate the global rank:} The complexity of ranking methods mainly depends on the complexity to compute centrality values of all nodes. Once the centrality values of all nodes are known then the global rank of the interested node can be computed in O(n) time by comparing its value with others. The local centrality measures are computed using local information and have very less complexity. But in the case of global centrality measures, the complexity is highly dependent on the given centrality measure and the network size. 

So, one can propose fast methods to compute the centrality rank. These methods either can approximate the centrality values or compute the centrality values of few selected nodes to estimate the rank of the interested node.

\item \textbf{Propose Methods to estimate the rank using local information:} The above discussed approach might require the entire network. In the case of large-scale networks, it is infeasible to collect the entire dataset and update it regularly. This creates a need to propose methods to estimate the rank using a small snapshot of the actual dataset that can be collected using graph sampling techniques. To collect the samples, a crawler is executed on the network, and it only requires the local information of the node to move to the next node. Thus, no global information of the network will be required.

\item \textbf{Propose methods to estimate the rank using the correlation of network properties:}
All real world networks possess some unique characteristics. We observe that the correlation of various centrality measures follows a similar pattern in scale-free real world networks; it will be discussed in section~\ref{structural_behavior}. It motivates us to exploit the network and node properties to estimate the global rank of a node. Global centrality measures have high computational complexity and their correlation with local properties can be used to propose fast approximation and ranking methods. For example, Wehmuth et al. \cite{wehmuth2012distributed} observed that the closeness centrality ranking is highly correlated with local neighborhood volume\footnote{Local neighborhood volume $Vol_h(u)$ of a node $u$ can be computed by adding the degree of all nodes which are reachable from the node $u$ in $h$ distance.} ranking, and this can be used to estimate the closeness rank of a node using its local information. To study the relationship between the node's rank and its properties, the machine learning models can be used. Once the model is trained, the complexity to compute the rank of a node is constant.

\end{itemize}

As per the best of our knowledge, there is no work in this direction. Various problems in network science like influence maximization, information spread, information diffusion, epidemic spread, are depending on the identification of the influential nodes \cite{kitsak2010identification, saxena2015understanding, dey2017literature}. For example, the influence spread in a network requires the identification of appropriate nodes to start the strategy. It is impractical to apply regular methods to identify influential nodes due to the large size and dynamic nature of the network. If we can propose efficient methods to identify influential nodes locally for different applications, it will be useful to control these dynamic phenomena.


\section{Methodologies and Progress So Far}

In this section, we discuss the methodologies used for the rank estimation and the obtained results.

\subsection{Using Structural Behavior of Centrality Measures}\label{structural_behavior}

We observe that each centrality measure follows a unique behavior in real world scale-free networks that can be exploited to estimate the rank.

\subsubsection{Degree Rank Estimation using Power Law Degree Distribution (PL Method)}

It is observed that real world scale-free networks follow power-law degree distribution \cite{barabasi1999emergence}. The probability $f(j)$ of a node having degree $j$ is given as $f(j) = cj^{-\gamma}$, where $c$ and $\gamma$ are constants for a network. Once we estimate the equation of the degree distribution, the number of nodes of each degree can be computed using the equation \cite{saxena2015rank}. The total number of nodes having degree $j$ can be computed as $n_j = n \cdot f(j)$, where $n_j$ represents total number of nodes having degree $j$ in network $G$. The degree rank of a node $u$ can be computed by adding the number of nodes having degree greater than the degree of node $u$ plus 1.

We prove that, in a scale-free network $G$ $(G \in \mathcal{G}(f))$, the parameters of power law degree distribution equation can be computed as, $c =  \frac{1- \gamma}{(d_{max})^{1-\gamma} - (d_{min})^{1-\gamma}}$ and $\gamma \approx 2 + \frac{d_{min}}{d_{avg}-d_{min}}$, where $d_{max}$, $d_{min}$ and $d_{avg}$ represent maximum, minimum and average degree of the network respectively, and $\mathcal{G}(f)$ is a set of networks following degree distribution $f$. After computing these parameters, the expected degree rank of a node $u$ can be computed as, $E[R_{G}(u)] \approx n \left( \frac{d_{max}^{1-\gamma} - (d_{u}+1)^{1-\gamma}}{ d_{max}^{1-\gamma} - d_{min}^{1-\gamma} } \right)  + 1$, where $n$ denotes the network size, and $d_u$ denotes the degree of node $u$ \cite{saxena2017degree, saxenaobserve}.

To apply the proposed method first we need to estimate network parameters like its size, minimum, maximum, and average degree using sampling method. In the simulation, the required parameters like average degree and network size are estimated using the methods proposed by Dasgupta et al. \cite{dasgupta2014estimating}, and Hardiman and Katzir \cite{hardiman2013estimating} respectively. The absolute error ($|Estimated \; Rank - Actual \; Rank|$) using power law (PL) method is shown in Figure 3 for BA (500,000 Nodes) \cite{barabasi1999emergence} and DBLP (317080 Nodes) \cite{yang2015defining} Network. This method does not give good results on real world networks as they do not follow the perfect power law. We further show that the variance in degree rank estimation using power law method increases from higher to lower degree nodes \cite{saxena2015estimating}. In the next section, we will discuss sampling-based methods to estimate the degree rank efficiently.

In random networks \cite{erd6s1960evolution}, degree distribution follows poisson distribution and the similar approach can be used to estimate the degree rank \cite{saxena2017degree}.

\subsubsection{Closeness Rank Estimation}

\begin{figure}[t]
\centering
\includegraphics[width=0.7\linewidth]{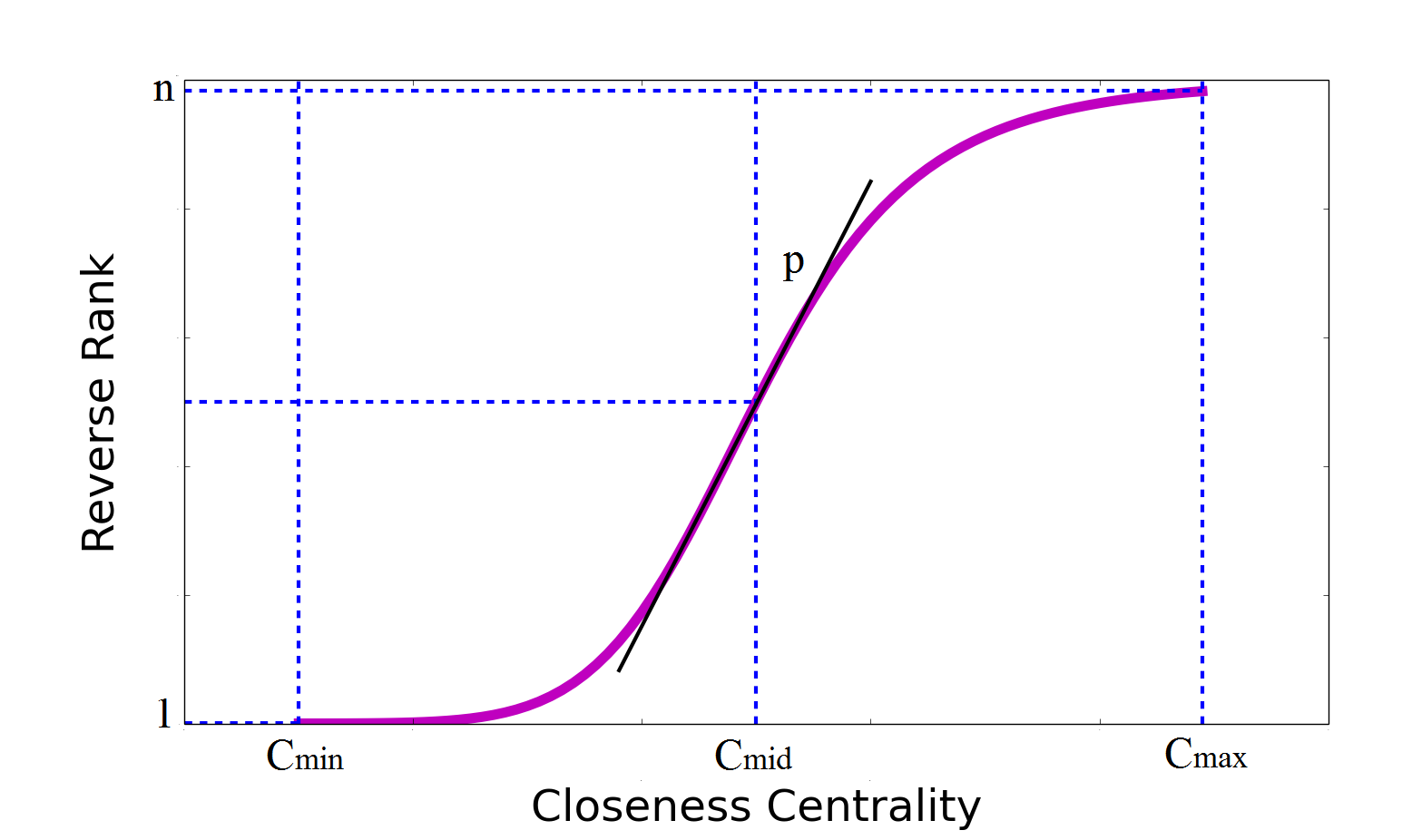}
\caption{Reverse Rank versus Closeness Centrality}
\label{fig:sorted}
\end{figure}

We study the structural behavior of closeness centrality and find that the reverse closeness centrality rank versus closeness centrality follows a sigmoid curve in social networks as shown in Figure~\ref{fig:sorted} \cite{saxena2017afaster, saxenafast}. In reverse closeness centrality ranking a node having the highest closeness centrality has the lowest rank and vice versa. We study this pattern and find that the 4 parameter logistic equation better fits the curve and can be used to estimate the rank. It is defined as,

\begin{center}
\label{reverserank}
$Reverse \; Rank(u) = n + \frac{1-n}{1+\left( \frac{C_C(u)}{c_{mid}}\right) ^p}$,
\end{center}
where $c_{mid}$ represents closeness centrality of the middle ranked node, and $p$ represents the slope of the logistic curve at the middle point.

The parameters of the sigmoid curve are estimated using other structural properties of the closeness centrality. We observe that the node having the highest degree either has the maximum closeness centrality or close to it. Similarly, the nodes having the minimum closeness centrality are farthest from the central node which has the maximum closeness centrality. These observations are used to estimate maximum and minimum closeness centrality that can be further used to estimate $c_{mid}$. In real world social networks, the slope of the logistic curve varies 11-15, and their average is used for the rank estimation.

\begin{figure}[]
     \begin{center}
         \subfigure[]{%
            \label{fig:first}
            \includegraphics[width=0.65\columnwidth]{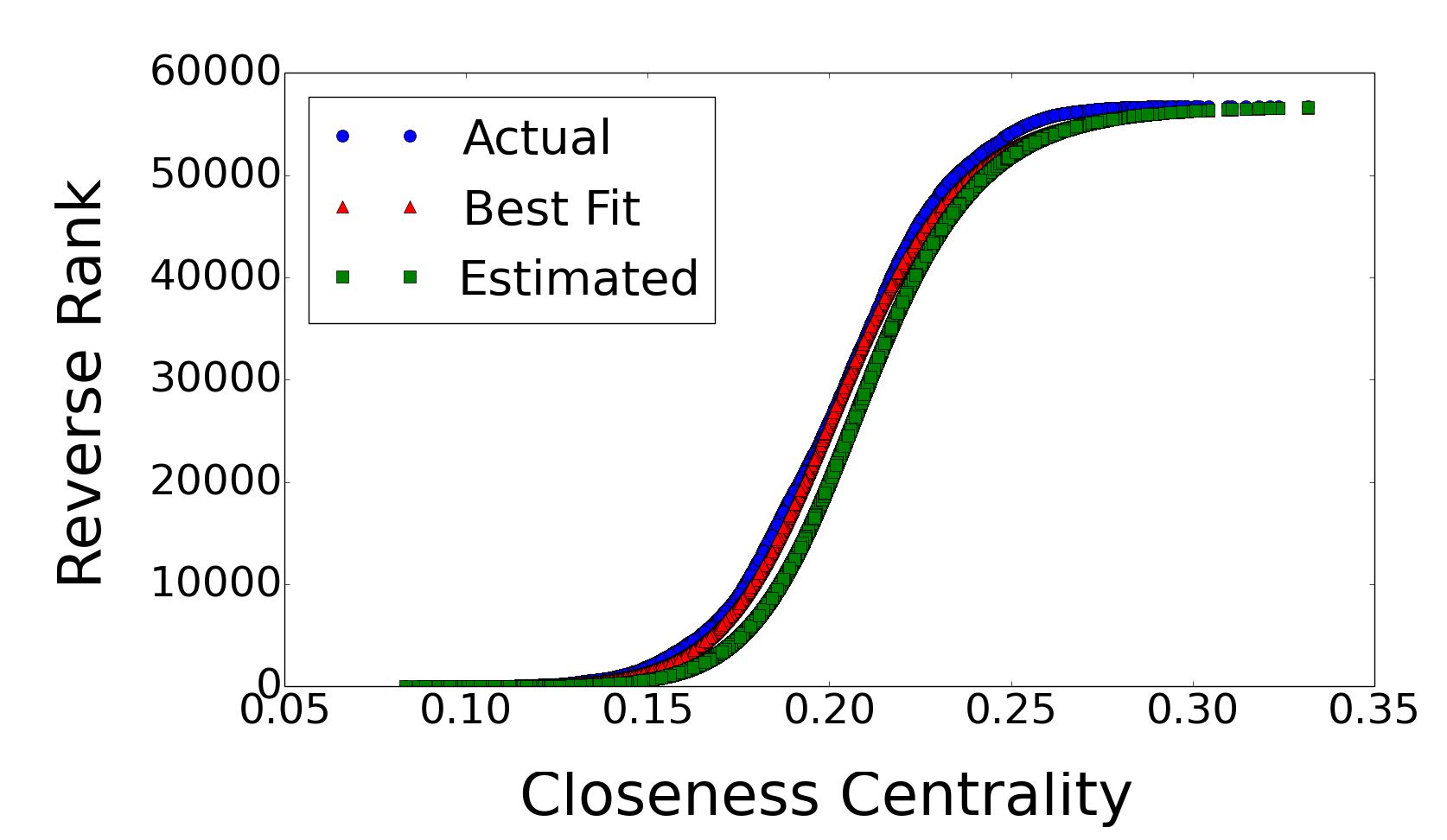}
        }\\%
        \subfigure[]{%
           \label{fig:second}
           \includegraphics[width=0.65\columnwidth]{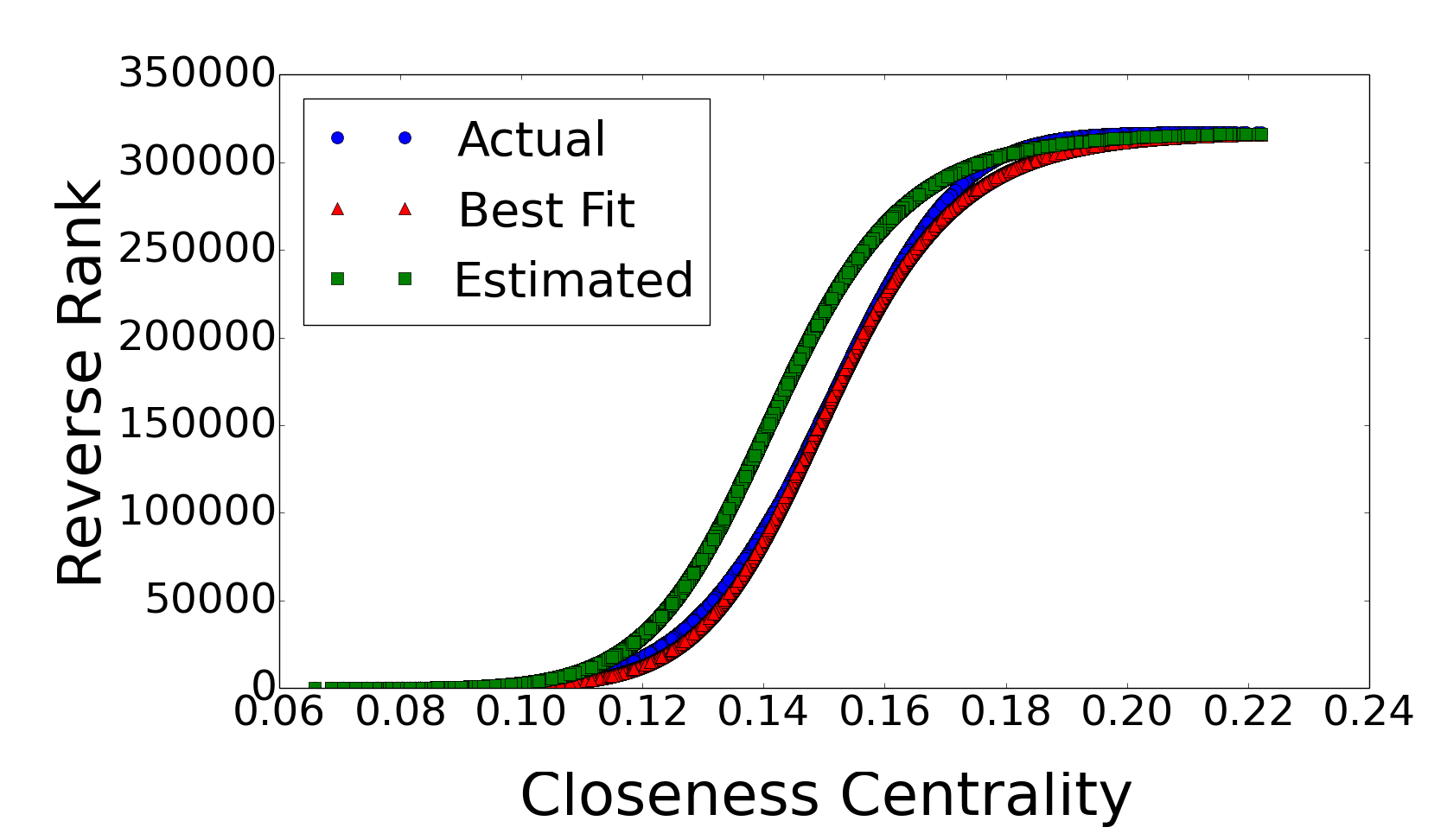}
        }
    \end{center}
    \caption{Reverese Rank versus Closeness Centrality for (a) Brightkite, and (b) DBLP Network}
   \label{sortclose1}
\end{figure}

In Figure~\ref{sortclose1}, the actual, best fit (4 parameter logistic curve), and the estimated curve for reverse rank versus closeness centrality is shown for Brightkite (56739 nodes) \cite{cho2011friendship} and DBLP (317080 nodes) \cite{yang2015defining} networks. Using this curve, the actual rank of a node $u$ can be estimated as, $Actual \; Rank(u) = n- Reverse \; Rank(u) +1$. This is a heuristic method and more details are available at \cite{saxena2017afaster}.


We further study this behavior for other centrality measures like eigenvector centrality, PageRank, triangle count, H-Index, and coreness, and observe that they all follow a unique pattern that can be used to estimate their ranking. 

\subsection{Using Sampling Techniques}

Due to the large size, it is impractical to collect the entire network to study its characteristics. Even if we collect the entire network to process a request, it will be of no use for further queries as the network would have been updated by that time. This has motivated researchers to use a small sampled dataset to study network characteristics like average degree \cite{dasgupta2014estimating}, average clustering coefficient \cite{hardiman2013estimating}, network size \cite{hardiman2013estimating}, and so on. We have also used various sampling techniques to estimate degree rank of a node that are discussed next.




\subsubsection{Degree Ranking using Uniform Sampling (US Method)}

In uniform sampling technique, the nodes are sampled uniformly at random. Once the samples are collected, the local rank of the \textit{interested node} is computed in the collected samples. This local rank is extrapolated to compute the global rank of the node. In a network $G$, using uniform samples, the degree rank of a node $u$ can be estimated as, $R_{est}(u) = \frac{n}{s}R_{local}(u)$, where $R_{local}(u)$ is the local rank of the node $u$ in the collected samples, $s$ is the sample size, and $n$ is the network size \cite{saxena2017degree}. 


In real world networks, uniform sampling is not feasible as the node ids and network size are not known in advance. In such scenarios, the sample can be collected using traversal based sampling techniques like random walk and its variations, BFT, DFT, etc. Next, we use metropolis-hastings random walk and random walk to estimate the degree rank.

\subsubsection{Degree Ranking using Metropolis-Hastings Random Walk (MH Method)}

In metropolis-hastings random walk, the probability distribution to move the crawler to the next node is modified so that the collected samples are equivalent to uniform samples \cite{metropolis1953equation}. 
The probability distribution function is defined as, 

$P_{u \rightarrow v} =\left\{\begin{matrix}
\frac{1}{d_u} \cdot min(1,\frac{d_u}{d_v}),  & if \; v \; is \; the \; neighbor \; of \; u, \\
1-\sum_{w \neq u}P_{u \rightarrow w}, & if \; v=u, \\
0, & otherwise.
\end{matrix}\right.$


After collecting the samples, the local degree rank of the interested node is computed in the collected samples. Then the global degree rank of the node can be estimated using the formula proposed for uniform sampling method.

\subsubsection{Degree Ranking using Random Walk (RW Method)}


In random walk samples, the probability of sampling a node is directly proportional to its degree. These samples can be converted to uniform samples using a new probability distribution function, where the probability of picking a node is inversely proportional to its degree $p(u) \propto 1/d_u$, known as re-weighted random walk sampling technique~\cite{hansen1943theory}.  After re-sampling, the global rank can be estimated using the local rank in the re-sampled dataset. The details are available at \cite{saxena2017degree}. 

\begin{figure}[]
     \begin{center}
         \subfigure[]{%
            \label{fig:first}
            \includegraphics[width=0.65\columnwidth]{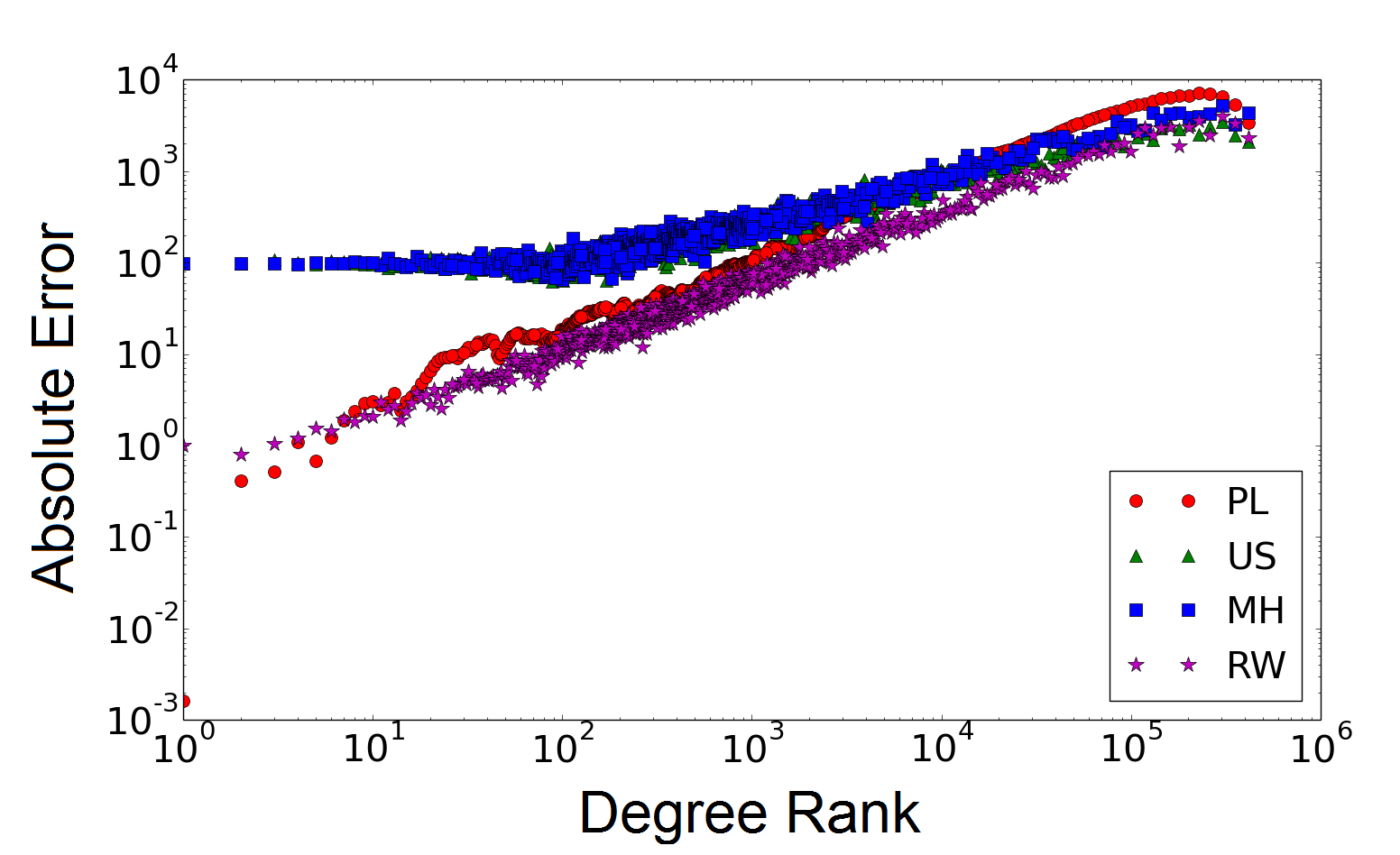}
        }\\%
        \subfigure[]{%
           \label{fig:second}
           \includegraphics[width=0.65\columnwidth]{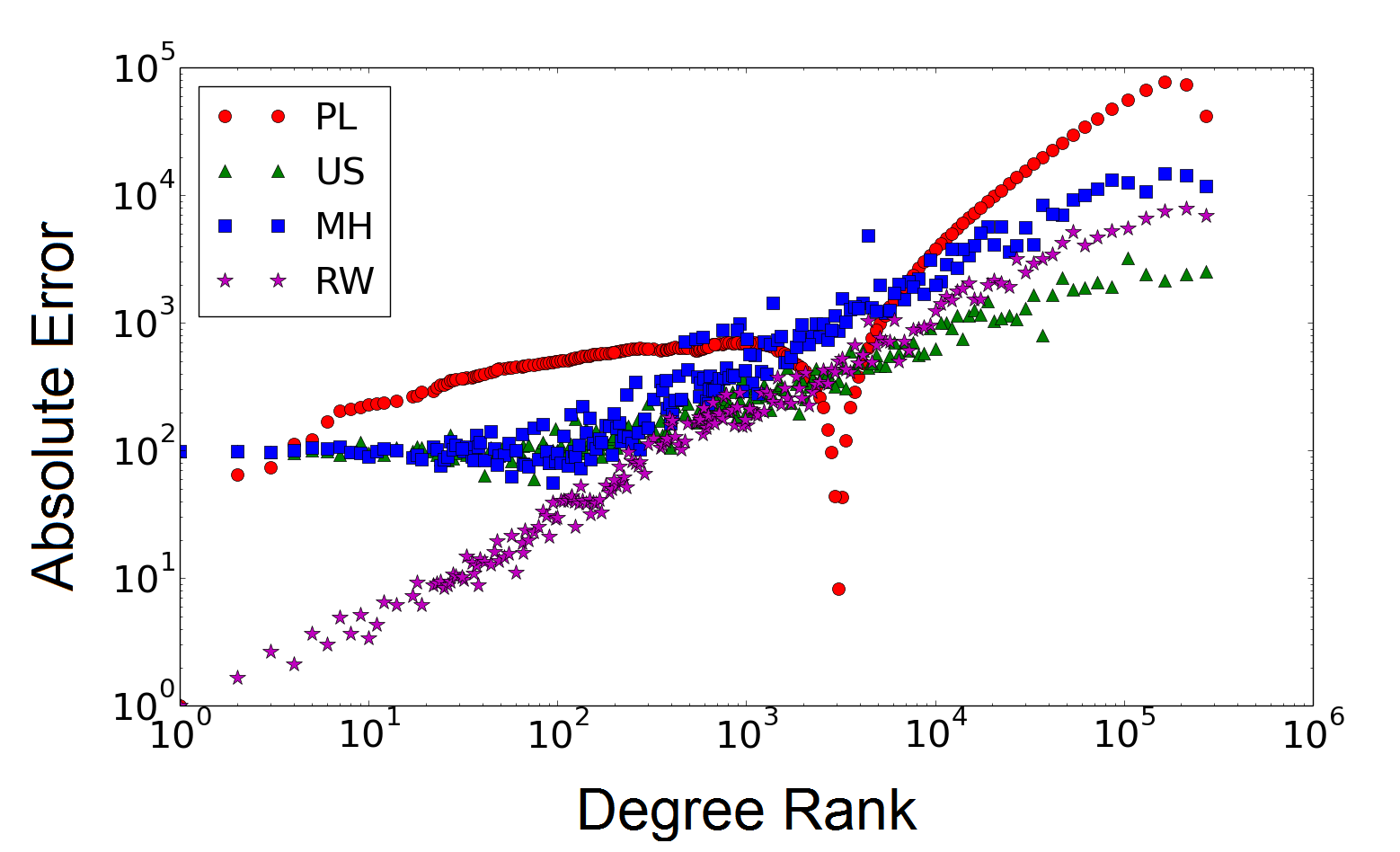}
        }
    \end{center}
    \caption{Absolute Error versus Degree Rank for (a) BA network (500000 nodes), and (b) DBLP network}
   \label{degerrfig}
\end{figure}

The results for degree ranking are shown in Figure~\ref{degerrfig}, where $x$-axis represents the degree rank and $y$-axis represents the absolute error ($|Estimated \; Rank - Actual \; Rank|$). In all sampling methods, the error is computed by taking the average of 20 iterations and each iteration collects $1\%$ samples. In all methods, the error increases from higher ranked nodes to lower ranked nodes. The results show that the RW method outperforms other methods, and, it can be efficiently used for real world networks. MH method gives more error than other sampling methods as the collected samples are not perfectly uniform samples for small sample size. We also observe that in all methods, the error decreases with an increase in the network size. The detailed results are explained here \cite{saxena2017degree}.

One can also propose new sampling techniques to estimate the rank efficiently. To apply the proposed strategy for global centrality measures, first, we need to approximate the centrality values using local information. Then the sampling methods can be used to predict the global rank of the node. These methods will be highly useful as the time complexity of global centrality ranking is very high.

\section{Conclusion}

In this project, we aim to propose fast and efficient methods to estimate global rank of the nodes based on different centrality measures. The proposed methods use the structural dependency of different centrality measures, sampling techniques, or machine learning models.

We have discussed four methods to estimate the degree rank without having the entire network. The power law method can be efficiently used to estimate the degree rank of a node in $O(1)$ time once the network parameters are estimated. In sampling-based methods, results show that the efficiency of random walk method is very close to uniform sampling method, and it can be used in real life applications.

We further discussed heuristic method to estimate closeness centrality rank of a node. The complexity of the proposed method is $O(m)$ (as it only computes the closeness centrality of three nodes, 1. closeness centrality of the interested node, 2. maximum and 3. minimum closeness centrality) that is a huge improvement over the classical method having the time complexity $O(n \cdot m)$.

The proposed methods are based on our understanding of the problem. One can also think of new directions to estimate the rank efficiently. In future, we would like to propose ranking methods for other kinds of networks like weighted networks, directed networks, multilayered networks, temporal networks, and so on.

\bibliographystyle{elsarticle-num}
\bibliography{/home/akrati/latex/mybib}

\pagebreak

\end{document}